\documentclass{sigchi}

\usepackage{balance}  
\usepackage{graphics} 
\usepackage{txfonts}
\usepackage{times}    
\usepackage[pdftex]{hyperref}
\usepackage{color}
\usepackage{textcomp}
\usepackage{booktabs}
\usepackage{ccicons}
\usepackage{todonotes}

\newcommand*{\eyeopen}{\includegraphics[scale=0.8]{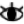}}
\newcommand*{\eyeclosed}{\includegraphics[scale=0.8]{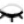}}
\newcommand*{\redlight}{\includegraphics[scale=0.8]{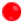}}
\newcommand*{\greenlight}{\includegraphics[scale=0.8]{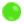}}
\newcommand*{\layerimport}{\includegraphics[scale=0.8]{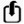}}
\newcommand*{\layerexport}{\includegraphics[scale=0.8]{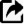}}
\newcommand*{\layermin}{\includegraphics[scale=0.8]{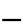}}
\newcommand*{\layermax}{\includegraphics[scale=0.8]{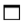}}
\newcommand*{\layerdup}{\includegraphics[scale=0.8]{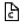}}
\newcommand{\ignore}[1]{}

\makeatletter
\def\url@leostyle{%
  \@ifundefined{selectfont}{\def\UrlFont{\sf}}{\def\UrlFont{\small\bf\ttfamily}}}
\makeatother
\def\pprw{8.5in}
\def\pprh{11in}

\definecolor{linkColor}{RGB}{6,125,233}
\hypersetup{%
  pdftitle={SIGCHI Conference Proceedings Format},
  pdfauthor={LaTeX},
  pdfkeywords={SIGCHI, proceedings, archival format},
  bookmarksnumbered,
  pdfstartview={FitH},
  colorlinks,
  citecolor=black,
  filecolor=black,
  linkcolor=black,
  urlcolor=linkColor,
  breaklinks=true,
}

\begin{document}

\title{Block Shelves for Visual Programming Languages}
\numberofauthors{3}
\author{
  \alignauthor{Sheng-yi Hsu\\
    \affaddr{National Chiao Tung University}\\
    \affaddr{Hsinchu,Taiwan}\\
    \email{syhsu@cs.nctu.edu.tw}}\\
   \alignauthor{Yuan-fu Lou\\
     \affaddr{National Chiao Tung University}\\
     \affaddr{Hsinchu, Taiwan}\\
     \email{yflou@cs.nctu.edu.tw}}\\
   \alignauthor{Chuen-tsai Sun\\
     \affaddr{National Chiao Tung University}\\
     \affaddr{Hsinchu, Taiwan}\\
     \email{ctsun@cs.nctu.edu.tw}}\\
}

\maketitle

\begin{abstract}
The blocks editor, such as the editor in Scratch \cite{resnick2009scratch}, is widely applied for visual programming languages (VPL) nowadays. Despite it's friendly for non-programmers, it exists three main limitations while displaying block codes: (1) the readability, (2) the program structure, and (3) the re-use. To cope with these issues, we introduce a novel formatting tool, \textbf{block shelves}, into the editor for organizing blocks. A user could utilize shelves to constitute a user-defined structure for the VPL projects. Based on the experiment results, block shelves improves the block code navigating and searching significantly. Besides, for achieving code re-use, users could use shelf export/import to share/re-use their block codes between projects in the file format of eXtensible Markup Language (xml.) All functions were demonstrated on MIT App inventor 2 \cite{pokress2013app}, while all modifications were made in Google Blockly.
%
%
%
%
%
%
%
%

\end{abstract}



\section{Introduction}

Visual programming languages (VPLs) are widely applied for reducing the programming barriers for non-programmers nowadays. Scratch \cite{resnick2009scratch}, MIT App Inventor \cite{pokress2013app}, Alice \cite{cooper2000alice}, etc., help the program learners significantly. Besides, many software tools provide VPL as an alternative programming environment for non-programmers, such as Orange \cite{JMLR:demsar13a} for data mining, Modkit \cite{millner2011modkit} and Ardublock for Arduino programming, LabVIEW \cite{elliott2007national} and Quite Universal Circuit Simulator (Qucs) \cite{JNM:JNM702} for circuit simulation, etc. Unlike text editors, VPL is intuitive and easy for non-programmers to program by drag-and-drop. Users first drag a block from a toolbox categorized with the block functions, and put the block in the blocks editor. With the concept of event-first programming \cite{Turbak:2014:EPA:2602724.2602739}, the orders or locations are irrelevant to the compiled result of a project/program.



However, according to Okerlund and Turbak \cite{okerlund2013preliminary}, a user has a fifty-fifty chance to deal with a project consisting more than 30 blocks while using MIT App Inventor. Undoubtedly, the more functions a user needs, the more blocks will be shown in the editor. From the perspective of personal usage, keeping tracks on all blocks becomes an issue when the blocks are accumulated to an extent. Meanwhile, in order to maintain a project, reminding oneself about all meanings of blocks are also difficult after a period of time; another remaining issue would be the tracing/understanding of blocks after opening others' projects for referencing or adapting. Another issues may cause problems when, for instance, a lecturer in a workshop/class needs the students to find a certain block in the blocks editor, or when the user want to share his/her blocks with others. In short, VPL effectively lower the barrier for program learning and developing, but unfortunately not for reading, tracing, and maintaining.

To address these challenges, we divide issues arisen from the current blocks editor into 3 categories: (1) the readability, (2) the program structure, and (3) the re-use of block codes. In order to cope with the issues, we introduce a novel formatting tool, \textbf{block shelves}, into the editor for organizing blocks in VPL projects. A block shelf in the editor is used for classifying blocks. Users could arrange the shelves from program functions, block usages, to the orders in an execution process, etc. Besides, every user could re-arrange the blocks various times without any deconstructions until finding the most suitable classification method. For more detailed explanations, the rest of the paper is organized as follow: background and related works, design overview, quantitative evaluation, and conclusion.   

%
%
%
%
%
%
%
%
%
%
%
%
%
\section{Background and related works}

Approaches in improving the code readability and the code usability have been researched for decades, but the techniques designed and discussed are mostly for the text-based programs, such as indentation \cite{Clifton:1978:TMS:953411.953415}, coding style \cite{Miara:1983:PIC:182.358437}, variable naming \cite{Teasley:1994:ENS:181031.181033}, modulation \cite{eick2001does}, and applying text context to improve the productivity \cite{Kersten:2006:UTC:1181775.1181777}. However, more and more projects today are block-based. The current blocks editor design provides 3 functions: block commenting, block collapsing and sorting by block category, and block duplicating to manage the readability issues, the structure, and the re-use of block codes. The details of the functions are explained in the following.    


\begin{itemize}
\item \textbf{Block commenting:} Adding comments to blocks is the most common method in the blocks editor to improve the code readability. Not only does it help others to understand your programs but it is also useful to keep one's own programs on track. Nevertheless, the current blocks editor doesn't provide a comment search function, so a user still needs to check blocks individually.

\item \textbf{Block collapsing and sorting by block category:} Block collapsing is originally designed for keeping the screen real-estate small; the collapsed blocks could be sorted with the block category. After sorting by block category, the editor would line up the collapsed blocks and show in a horizontal or vertical way. Though block collapsing and sorting by block category makes the block layout succinct and clean, the block layout still lacks a structure or organized system for users to trace the codes.    
\item \textbf{Block duplicating:} Block duplicating is an operation for users to copy and paste a group of selected blocks. However, this operation is only for single project at the moment; therefore, if a user wants to adapt other projects, he/she has to "imitate"(re-build) the blocks from the referenced project.
\end{itemize}


\section{Design overview}
The design of block shelves provides a user-defined structure for shelving blocks in VPL projects. We chose MIT App Inventor 2 (AI2) to demonstrate the block shelves. In MIT AI2, the blocks editor is a Javascript-based Google Blockly combined with Google App Engine. Hence all modifications for demonstrating shelves are made on Google Blockly. 

\begin{figure}
\begin{minipage}[t]{0.5\linewidth}
    \centering
    \includegraphics[width=0.9\textwidth]{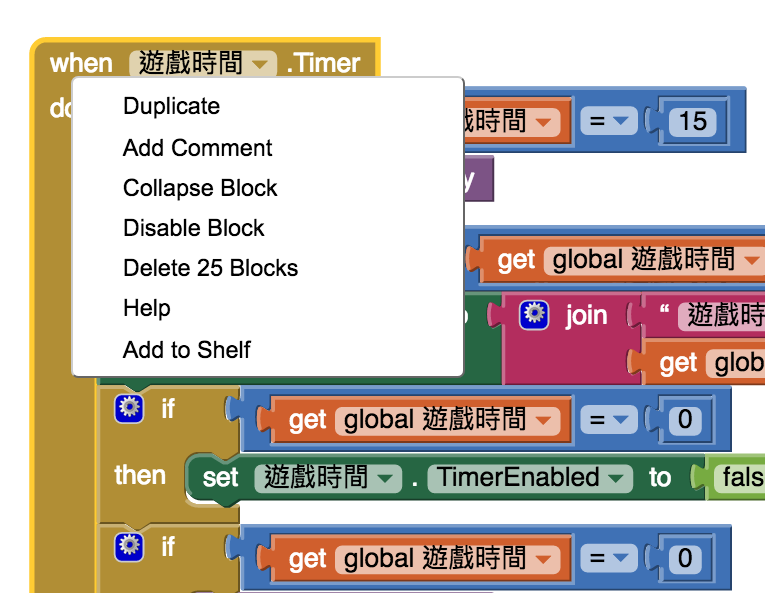}
    \caption{The menu for adding a shelf.}
    \label{fig:layer}
\end{minipage}
\hspace{0.1cm}
\begin{minipage}[t]{0.4\linewidth} 
    \centering
 \includegraphics[width=0.5\textwidth]{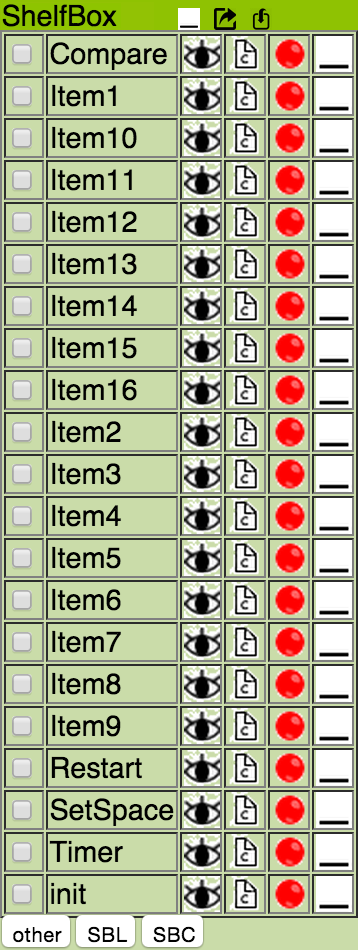}
    \caption{ShelfBox: the toolbox of shelves.}
    \label{fig:toolbox}
\end{minipage}
\end{figure}



    

Adding blocks to a shelf is same with using other block functions in AI2. The menu is shown in figure \ref{fig:layer}. A user could use the right mouse key to add a group of blocks to a shelf in a project. The added shelf is shown in the shelf toolbox, ShelfBox (figure \ref{fig:toolbox}). The main function designs for shelves are shelf visibility, shelve minimization/maximization, shelf activation/deactivation, shelf duplication, and shelf exportation/importation. The details of each design are described in the following.

\begin{itemize}
\item \textbf{Shelf visibility (\eyeopen{} / \eyeclosed{}):} Users could utilize this function to hide or show the blocks in the selected shelf in the blocks editor. This design not only saves the screen space efficiently but controls the numbers of blocks shown in the block editor. 

\item \textbf{Shelf minimize/maximize (\layermin{} / \layermax{}):} This function is adapted from block collapsing/expanding. By collapsing or expanding all blocks in selected shelves, a user controls the numbers of blocks shown in the block editor effectively. However unlike shelf visibility, one still sees a minimized block in the editor after applying this function. 

\item \textbf{Shelf activate/deactivate (\redlight{} / \greenlight{}):} Shelf activation/deactivation is used for activating or deactivating all block functions in the selected shelves. Unlike shelf minimize/maximize, this design is used for tracing codes by checking the block functions.

\item \textbf{Layer duplicate (\layerdup{}):} This design is similar to block duplicating; the only difference is that this function duplicates all blocks in a selected shelf, which is useful when creating similar programs or testing functions without changing the original programs.

\item \textbf{Shelf export/import (\layerexport{} / \layerimport{}):} This design is used for sharing blocks between different projects. In order for better code-reuse, a user utilizes this design to share the selected shelves in the file format of eXtensible Markup Language (xml) with other users.
\end{itemize}

\section{QUANTITATIVE EVALUATION }
Our main focus are code-reading and -navigating; therefore, we first define the twin activities of reading and navigating as code understanding \cite{bragdon2010code}, using block searching time and block reading time to measure the differences between the block editors with and without block shelves. In our research framework, the sample was divided into one experimental (App inventor 2 with block shelves) and one control group (original App inventor 2). This evaluation investigates the following hypotheses: (1) whether block shelves improve code understanding, (2) whether block shelves improve code re-use, and (3) whether block shelves improves the searching time of locating a certain block. 

\subsection{Experimental Design and Procedure}
A total of 60 graduate and undergraduate students (mean age 23, max age 38, min age 18) were recruited from National Chiao Tung University or National Tsing Hua University. They never use VPL but ever took the courses of computer introduction and programming. All participants are randomly divided into either experimental or control group, 30 students for each group. The experiment includes 3 stages: (1) tutorial, (2) pre-test, and (3) post-test. Both groups joined the tutorial stage and pre-test. In the post-test stage, the experimental and control groups are separated: the experimental group took a introductory tutorial for block shelves before using App inventor 2 with shelves, while the control group used the original App inventor 2. Meanwhile, all trials are conducted with the Chrome browser and recorded with oCam v23.0 (video capture software.) The operating system is Windows 7, and the spec of computers for trials is Intel Core i5 2400 @ 3.10GHz (CPU) with 4GB of ram, a 22" monitor running at 1920x1080@60Hz, and a NVIDIA GeForce GT 630.  

\begin{figure}
\begin{minipage}[t]{0.45\linewidth}
    \centering
    \includegraphics[width=0.8\textwidth]{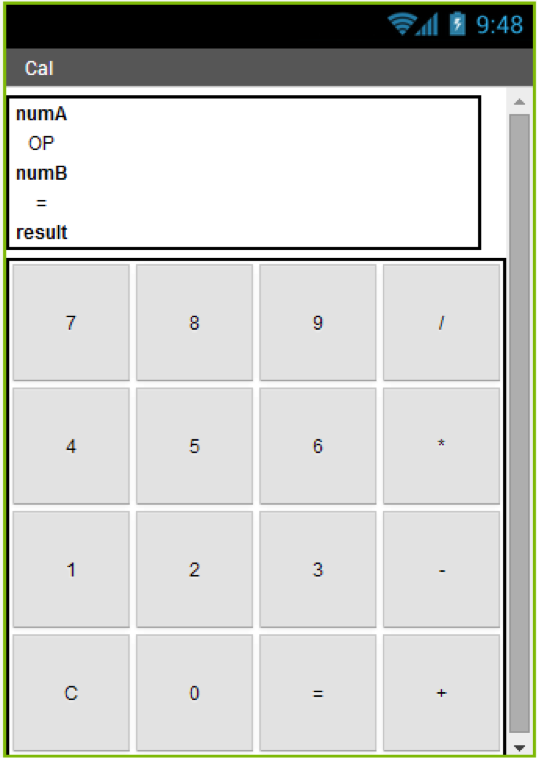}
    \caption{Project for pre-test: Calculator.}
    \label{fig:calculator}
\end{minipage}
\hspace{0.2cm}
\begin{minipage}[t]{0.45\linewidth}
    \centering
    \includegraphics[width=0.8\textwidth]{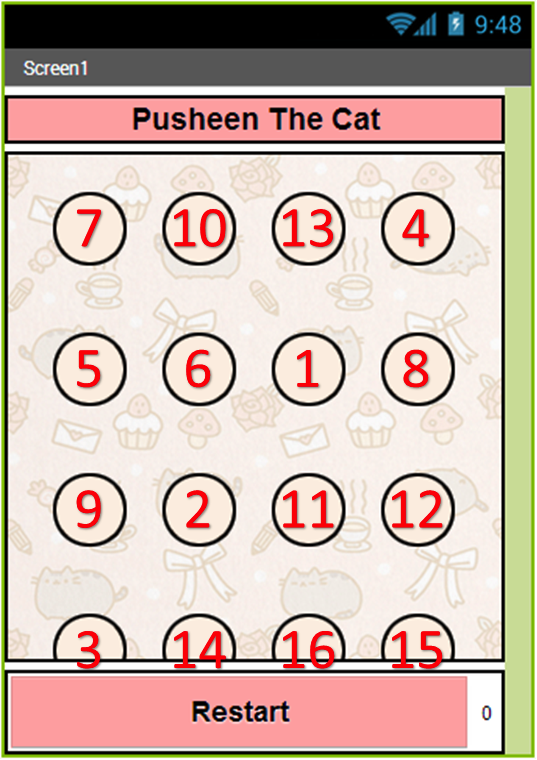}
    \caption{Project for post-test: Pusheen The Cat}
    \label{fig:pusheen}
\end{minipage}
\hspace{0.1cm}
\end{figure}

In the tutorial stage (40 minutes), we introduced the Android app front-end design and the block code design. The tutorial project includes 197 blocks and 6 different screens to show users how to add a gadget to the mobile phone interface and use block codes to control the gadgets on the interface, including how to add a button, a canvas, a ball, a timer, a graph, and a text tag. Besides, the pre-test adopted calculator (figure \ref{fig:calculator}), which is designed for both groups to test whether abilities between the two groups are identical; In the post-test, Pusheen the cat (figure \ref{fig:pusheen}) is adopted for testing whether the shelves improves the user performance. Pusheen the cat is a memory game, which requires a player to remember the location of each number. Every time a player needs to pick two buttons to press and a player earns two points when the sum of the two numbers of the selected buttons is 17. To complete a game a player has to achieve 16 points, and the number in the lower-right corner of the screen shows how long a player used. The tasks in each project are listed in the table \ref{tab:tasks}.

In Calculator (fig. \ref{fig:calculator}), task 1 and task 2 ask users to add the functions of division and number 9 by referencing other arithmetic operations and numbers. These two tasks are designed for checking the conditions of code re-use in the original blocks editor. The rest tasks (task 3, 4, and 5) are used for observing the code understanding (code reading and code navigating) because users need to locate the target block and then to correct the target block in these three tasks. On the other hand, the block shelves used by the experiment group in Pusheen The Cat (fig. \ref{fig:pusheen}), are constructed first with the buttons, as shown in figure \ref{fig:toolbox}. In Task 1 and task 3, users are asked to add/correct block codes by referencing other items. Through the two tasks we want to see if block shelves are helpful for code re-use. Task 2, task 4, and task 5 are used for checking whether layers help code understanding since users are asked to locate the target block to fix. 


\begin{table*}
\centering
\caption{Tasks in Calculator and Pusheen The Cat}
\begin{tabular}{ccc}
\centering
    Task & Calculator (Pre-test) & Pusheen The Cat (Post-test) \\
    \hline
    1. & Add the function of button $\langle$/$\rangle$ & Add the function of button $\langle$item 7$\rangle$ \\
    2. & Add button $\langle$9$\rangle$ & Correct the response when choosing two identical photos \\
    3. & Correct the function of button $\langle$+$\rangle$ & Add the display of a cat's photo after hitting $\langle$item 5$\rangle$ \\
    4. & Correct the function of button $\langle$5$\rangle$ & Correct the timer reset after hitting button $\langle$Restart$\rangle$ \\
    5. & Correct the function of button $\langle$=$\rangle$ & Correct the text alert when game is over or finished \\
    \hline
\end{tabular}
\label{tab:tasks}
\end{table*}


\subsection{Results and Analysis}
\begin{figure}
\begin{minipage}[t]{\linewidth}
    \centering
    \includegraphics[scale=0.7]{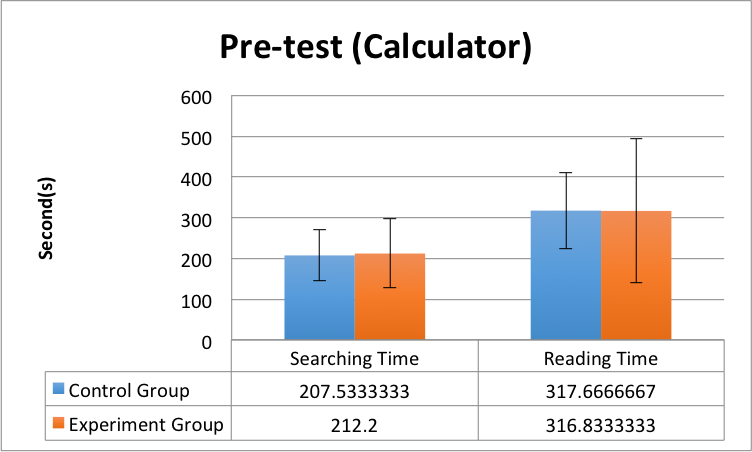}
    \caption{The pre-test results of the searching time and the reading time.}
    \label{fig:pre_test}
\end{minipage}
\hspace{0.1cm}
\begin{minipage}[t]{\linewidth} 
    \centering
 \includegraphics[scale=0.7]{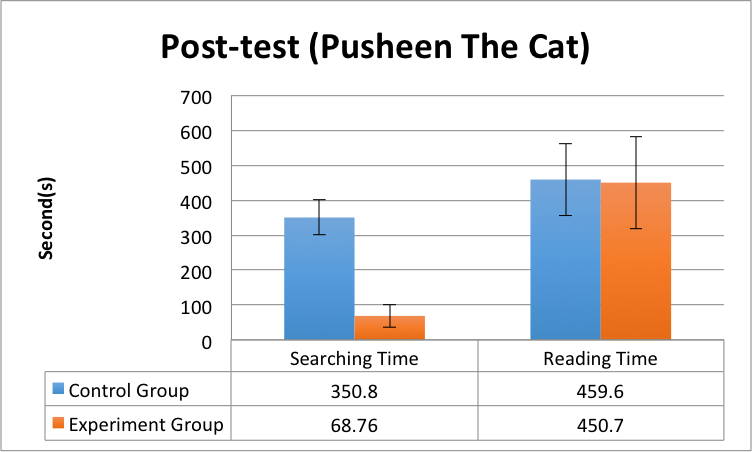}
    \caption{The post-test results of the searching time and the reading time.}
    \label{fig:post_test}
\end{minipage}
\end{figure}

We measured every participant's reading and searching time while doing Calculator and Pusheen The Cat. The searching time refers to the time a participant used for locating the target block after reading a task, and the reading time indicates the time a participant spent to find the bug after locating the target block. The averages and the standard deviations of the pre-test and the post-test are shown in figure \ref{fig:pre_test} and figure \ref{fig:post_test}. 

We applied \textit{Levene} test to check if the variances of the two groups are equivalent. In the pre-test, Calculator (table \ref{tab:tasks}), we first check the searching abilities between the experimental group and the control group: the differences between the two groups are not significant ($P(T\le\textit{t})=0.750>.05$) based on the assumption of equal variance in \textit{t}-test ($P(F\le\textit{f})=0.118>.05$). This test result also implies the abilities between the two groups were similar in the searching abilities. Further we checked the reading abilities between the two groups. The result of \textit{t}-test assuming equal variance ($P(F\le\textit{f})=0.306 >.05$) shows that the reading abilities between the two groups are also similar ($P(T\le\textit{t})=0.974 >.05$). Combining the results mentioned above we concluded that the reading and searching abilities of block codes between the two groups are similar. 

After confirming the ability levels between the two groups, we proceeded to the post-test. For the searching time, we found that the experimental group used significantly less time than the control group ($P(T\le\textit{t})=4.727E-19<.05$) based on \textit{t}-test assumed unequal variance (P(F<\textit{f})=6.897E-07 <.05). However, the two groups were not significantly different ($P(T\le\textit{t})=0.826>.05$) in the time spent on reading according to \textit{t}-test assumed equal variance ($P(F\le\textit{f})=0.062 >.05$).

The post-test results showed that the block shelves help users navigate and locate the target blocks efficiently. However, as for the reading of block codes, the introduce of block shelves doesn't improve significantly because the reading time is highly correlated with the extent of a user's familiarity of block codes. All participants in our experiment were their first time to use the block editor. Hence the block shelves provides users an formatting tool for improving navigating, searching, and tracing in block-based projects. 


\section{Conclusion}

We introduce \textbf{block shelves}, a novel formatting tool, for organizing the block-based projects. The experiment results suggest that the block editor with block shelves helps users navigate and search block codes, which improves the  searching time significantly. By adopting shelves users could organize the block codes with their own arrangements. Besides, we also designed five functions for shelves: (1) shelf visibility, (2) shelf minimize/maximize, (3) shelf activate/deactivate, (4) shelf duplication, and (5) shelf export/import. The first four functions are adapted from the original block functions, and the last function is specially designed for cross-project sharing/re-use. Combining the above points, block shelves provides a useful shelving system for the current VPL projects, breaking through the conventional structure and offers a brand new interface for users.       


\bibliographystyle{SIGCHI-Reference-Format}
\bibliography{reference}

\end{document}